%Paper: hep-ph/9302249
%From: Sylvie Zaffanella <zaf@amoco.saclay.cea.fr>
%Date: 11 Feb 93 14:36:22+0100

% Following Postscript files are included: fig1.ps fig2.ps fig3.ps fig4.ps
%%fig5.ps fig6.ps
% (Files are separated by lines %%%%%%%%...)
% Plain TeX file

\font\huitrm=cmr8
\magnification 1200
\input epsf
\def\epsfsize#1#2{\ifnum#1>\hsize\hsize\else#1\fi}
\hoffset=1truecm
\hsize=36truepc
\vsize=51truepc
\voffset=0truecm
\baselineskip=14truept
\parindent=1 truecm

\def\caption#1{{\leftskip=1.3truecm

\rightskip=1.3truecm
\huitrm
\noindent
\baselineskip=12truept
 #1\par}}
\def\adresses#1{{
\huitrm
\noindent
\baselineskip=12truept
 #1\par}}

\def\half{{1\over 2}}

\def \={\ =\ }
\def \exp#1{{\rm exp}\left[#1\right]}
\def \ref#1{$^{[#1]}$}
\def \moy#1{\langle{#1}\rangle}

\def \bigf{I\!\! F}

\def\as{\alpha_s}
\def\epm{$\rm e^+$-$\rm e^-$}
\def\g0{\gamma_0}

\def\frac#1#2{{#1\over #2}}

\def\equat#1{\eqno {\rm #1}}

\def\figur#1{\null$$\epsfbox{#1}$$}
\def\partder#1{{\partial\over \partial #1}}
\def\Form#1{[$\>{\rm #1}\>$]}
\line{\vbox{\halign{\hfill#\hfill\cr
Nice INLN 93/1\cr}}
\hfill\vbox{\halign{\hfill#\hfill\cr
January 1993\cr}}}

\noindent Saclay Spht/93-011
\vglue3cm
\centerline {\bf ANGULAR INTERMITTENCY IN QCD JETS}
\vskip2truecm

\centerline { Ph. Brax$^\#$, J.-L.
Meunier$^*$,
and R. Peschanski$^\#$ }
\bigskip
\vskip 2truecm
\centerline{ ABSTRACT}
\medskip
Using two methods, via fluctuations and correlations,
an analytical formula is derived for the factorial multiplicity
moments in a QCD jet at the Double Leading Logarithm accuracy. The
resulting
self-similar dependence on the solid-angle cell size is
characteristic
of an intermittency behaviour in angular variables. The intermittency
indices depend on the diffusion angle  through the running
of
$\alpha_{s}$. Physical features of jet fluctuations
such as collimation at large angles and saturation  at small angles
are well described in the perturbative framework.
A parameter-free prediction of angular intermittency is proposed for
${\cal Z}^0$ decays into hadrons, assuming hadron-parton duality.

\vskip 3truecm

\vfill

\adresses{\noindent * {Institut Non Lin\'eaire de Nice, Universit\'e
de Nice-Sophia
Antipolis}, {Parc Valrose, 06108 Nice Cedex 2, France}: Unit\'e Mixte
de
Recherche du CNRS, UMR 129, }

\adresses{{$^\#$ Service de Physique Th\'eorique}, {Centre d'Etudes
de Saclay, F-91191 Gif sur Yvette cedex, France}:
Laboratoire de la Direction des Sciences de la Mati\`ere du
Commissariat \`a l'Energie Atomique.}

\eject\vglue 2.9truecm
{\bf I Introduction}
\bigskip
The observation of multiplicity fluctuations in multiparticle
production reactions at high energy is quite old \ref{1}, but it is
only relatively recently\ref{2} that a quantitative method of
analysis has been proposed and its results discussed in terms of the
supposed reaction dynamics. Interestingly enough, a set of non
trivial dynamical (by opposition to statistical) fluctuations has
emerged, with the intriguing possibility\ref{2,3} of a self-similar
scaling behaviour as a function of a small cell size in momentum
space,
called "intermittency" by analogy with hydrodynamics. However, the
 system at stake being ultrarelativistic and involving Quantum
Physics, is very far from the classical areas where
intermittent behaviour have been studied; Moreover, in Particle
Physics, it is not easy to identify the possible origin of
intermittency and more conventional models has been proposed as an
alternative\ref{4}. It is thus of importance to investigate this
problem in the framework of the standard theory of strong
interactions - Quantum Chromodynamics (QCD) - and in a case where it
is
known in principle to be calculable, namely in the weak-coupling
perturbative regime of jet physics. This is the goal of the present
paper.

The method designed to study multiplicity fluctuations from event to
event in a given phase-space cell (to be defined later on) makes use
of the so-called scaled factorial moments expressed as follows:
$$
{\cal
F}_q(\Delta)={\moy{n(n-1)\dots(n-q+1)}_\Delta\over\moy{n}_\Delta^q},
\equat{I-1}
$$
where $n$ is the particle multiplicity in a phase space cell of size
$\Delta$, in which multiplicity fluctuations are expected.
 Intermittency is characterized by a scale-invariant behaviour
of the scaled factorial moments, namely
$$
{\cal F}_q(\Delta)\propto \Delta^{-f_q}\equat{I-2}
$$
where the constants $\{f_q\}$, usually called intermittency indices,
play the role of anomalous dimensions of the fluctuation pattern.
Indeed, the intermittency indices can in principle be related to the
Renyi dimensions ${\cal D}_q$ characterizing a set of fluctuations.
It reads\ref{5}
$$
{\cal D}_q=1-{f_q\over q-1}\equat{I-3}
$$
where ${\cal D}_q$ measures the defect of phase-space dimensionality
due to
intermittency (the dimension $1$
in \Form{I-3} corresponds to an uniform distribution in the
$\Delta$ cells for all events).

It is to be noted that the same relations \Form{I-1,2,3} can be
quite
equivalently interpreted and discussed in terms of
correlations\ref{6}. Indeed, on has the following relation:
$$
{\cal F}_q(\Delta)\propto \int\dots\int_\Delta\prod_1^q d\omega_i
\rho(\omega_1,\dots,\omega_q)
$$
where $\rho$ is the differential multiplicity density distribution
for $q$ particles in the phase space $\Delta$ integrated over the
variables
$\omega_i$.  In
particular, the existence of an intermittent
behaviour \Form{I-2}, can be traced back to the existence of a
singularity
in $\rho$ at very short momentum distances\ref{7} and thus in the
corresponding correlation functions\ref{6}. In fact, these two
different "languages", fluctuations and
correlations, though equivalent, lead to two different calculation
methods of the factorial moments in the perturbative QCD framework,
as we shall see further on. This will serve as a non trivial
check of the calculations, as the final result is obtained in
 both ways.

A behaviour compatible with \Form{I-2} has been found in various
reactions
analyzed in terms of the rapidity $y$, azimuth angle $\varphi$ and/or
transverse momentum $P_\perp$ of the produced particles around the
jet
reference axis\ref{8}, giving rise to a lot of phenomenological
discussions and models of various types\ref{9}. However, the
observation of an intermittent behaviour in \epm anihilation into
hadrons\ref{10} and especially in the ${\cal Z}^0$ hadronic decays at
LEP\ref{11} has raised a particular interest. Beside the quality of
high-precision, high-statistic events allowing the non-trivial
extraction of the factorial moments from data, it offers    the
possibility
of a comparison with perturbative QCD
predictions, either "indirectly" through Monte Carlo simulations
including full hadronisation models or "directly", at the level of
partonic shower and assuming parton-hadron duality\ref{12}.

Indeed, the evidence that Monte-Carlo models of \epm
annihilation into jets could naturally lead to intermittent patterns
has
been numerically obtained in Ref.[12], but the hadronization scales
imply a "saturation effect", namely a stabilisation of factorial
moments beyond some small cell-size in rapidity. As a matter of fact
at LEP
energy,
Monte-Carlo models based on Perturbative QCD plus an hadronization
model give
a satisfactory description\ref{11} of 1, 2 and 3-dimensional factorial
moments, with some non-negligeable, but limited      discrepancies
when various experimental cuts are performed. However, here also, the
role of hadronization in the final result can be important, or at
least remains ambiguous. Indeed, two different Monte Carlo's
procedures,
the "Parton Shower Models" and the
"Matrix Element Models" give quite similar results\ref{11}, while the former
admits a rather extended parton cascading and the second is limited to
at most 4 final partons before hadronisation takes place. In this case,
almost all the effect is encoded in the hadronization model.

On the theoretical side of the question, it is possible to estimate
factorial moments at the parton level using  the
leading log approximation of the perturbative expansion.
For instance, evidence for an intermittent behaviour was
numerically obtained from a simulation of a gluon cascade\ref{13},
using a simplified version of a QCD jet. The results can be compared
with
data by assuming the LPHD\ref{14} (local parton-hadron duality
hypothesis).
Then, a quite systematic
study
has been analytically performed in the framework of the
so-called "dipole approximation"  of the quark-gluon
radiation
allowing a determination\ref{15} of fractal dimensions related to
gluon
cascading. Note that these dimensions are considered in Ref. [15] as
being
different from intermittency anomalous dimensions, we will later come
back to this interesting discussion.
Two more recent papers, one using planar diagrams\ref{16}, and one in
the framework of multiparton correlations\ref{17} will also be
discussed in comparison with our methods and results.

Our aim is to obtain the most general Perturbative QCD prediction at
the
double leading log approximation (DLLA) for the pattern of
fluctuations/corre-
lations within a parton jet. In the analysis based on fluctuations,
one
will be inspired by some methods used in the paper of ref. [16]. For
the
correlation analysis, the method is borrowed from the calculation of
energy-multiplicity correlators, which already led Y. Dokshitzer and
S. Troyan
to the discussion of factorial moments \ref{18} and have been
recently developped in a different context, see
\ref{19}.

The key points we want to address, using
the present state of the art\ref{20},
can be summarized in a series of questions:

\hskip 1truecm{\it i)} Is Perturbative QCD intrinsically intermittent
?

\hskip 1truecm{\it ii)} In which variables this property of
fluctuations shows up

\hskip 1truecm{\it iii)}  What are the anomalous dimensions
characterizing the intermittency behaviour?

\hskip 1truecm{\it iv)} How to perform the phenomenological   analysis   of the
DLLA predictions?

\hskip 1truecm{\it v)} Why intermittency appears in Perturbative QCD
calculations?

As we shall now see in the derivation, the main result of our study
is
that                the question {\it i)} is investigated in detail
and
answered positively; From that study, one also gets a precise
 answer to questions {\it ii)-v)}. This is the
subject of the present
paper, which is organized in the following way:
In  section II, we use  a first method - the {\it fluctuation}
approach - to get a first hint to the calculation of the factorial
moments and the emergence of angular variables in the structuration
of fluctuations. However, an ambiguity remains about the energy
variable, which in section III is resolved by the method of
{\it correlations}
 while confirming the validity of the
first results. The
 fourth section is devoted to the description of the resulting
intermittent behaviour and to a discussion of the important
difference between fixed and running coupling constant which
leads to
QCD scaling violation effects of a new nature ("saturation effects").
The
resulting "angular intermittency" property is discussed both for its
phenomenological implications and its comparison with previous
approaches.
Finally, in section V,
we propose  some comments on the roots of the intermittent behaviour
in a
quantum field theory like QCD and discuss open
problems
and suggestions for further study.

\bigskip

{\bf II Intermittency from fluctuations in a QCD jet}
\bigskip
Let us start from the well-known multiplicity generating
function\ref{20} of one gluon jet with opening angle $\Theta_0$ and
energy $E$, ${\cal Z}(E\Theta_0,v)$:
$$
\eqalign{ &{\cal Z}(E\Theta_0,0)=1\cr
&{\partial {\cal Z}\over \partial v}\big|_{v=0}=\moy{n}=\bigf_1\cr
&{\partial^q {\cal Z}\over \partial
v^q}\big|_{v=0}=\moy{n(n-1)(\dots)(n-q+1)}=\bigf_q\cr}\equat{II-1}$$
where $\bigf_q$, by difference with \Form{I-1},
are for non-normalized and {\it global}
multiplicity moments. Note that the generating function depends kinematically
on the product $E\Theta_0$, i.e. the highest parton transverse
momentum\ref{20}.

 ${\cal Z}(E\Theta_0,v)$ obeys the following evolution equation
\ref{21}~:
$$
{\partial{\cal Z}\over \partial \ln \Theta_0}\equiv \dot{\cal
Z}={1\over 2}\int_{z_0}^{1} {\alpha_s\over 2 \pi}\Phi_A(z)\left[{\cal
Z}(Ez\Theta_0,v){\cal Z}(E(1-z)\Theta_0,v)-{\cal
Z}(E\Theta_0,v)\right]dz\equat{II-2}
$$
where $\Phi_A(z)=4C_A\left({1\over z}+{1\over 1-z}+ ... \right)$
$C_A$
is the gluon color factor, and $z_0 =\mu /E\Theta_0$ the cut-off from
the non-perturbative region. This equation is given in terms of an {\it
energy integral} and  seems a priori different from        the
{\it   multiplicity density integral}  typical  of an intermittent
Random
Fragmentation Model\ref{22} whose evolution  is however also governed
by
a quadratic non-linear equation (see farther, Eq.\Form{ II-8,9}).
Up
 to non leading terms the equation \Form{II-2} can be
simplified to :
$$
\dot{\cal Z}\simeq{\cal Z}\int_{z_0}^1\gamma_0^2{dz\over z}\left[{\cal
Z}(Ez\Theta_0,v)-1\right]\equat{II-3}
$$
with $\gamma_0^2=4C_A{\alpha_s/ 2\pi}$.

Let us first consider the simplified case of a fixed coupling
constant $\as$.
The first-moment equation can be integrated from the smallest angle
possible, $\Theta_\mu=\mu/E$, (where $\mu$ is some infrared cut-off
of the order of the  hadronic scale), up to $ \Theta_0$:
$$
\moy{n}\equiv\bigf_1\propto \exp{\gamma_0\ln{E\Theta_0\over
\mu}}\equat{II-4}
$$
which represents the well-known total multiplicity of gluons in a jet
of angular aperture $\Theta_0$ in the fixed-coupling regime. Then,
the other moments equations can be obtained from \Form{II-3} as :
$$
\dot{\bigf_q}=\sum_1^{q}C^p_q\bigf_{q-p}\int_{z_0}^\infty{dz\over
z}\gamma_0^2\bigf_p(zE,\Theta_0)\equat{II-5}
$$
where the $C_q^p$ are the binomial coefficients. As is well
known\ref{20} such distributions follow asymptotically the KNO
scaling law, namely$$
\bigf_p=c_p\bigf_1^p\equat{II-6}
$$
where $c_p$ are known constants. Using \Form{II - 4} and \Form{II -
6}, the integral in \Form{II - 5} can be done explicitely, and the
equations takes the following form:
$$
\dot{\bigf_q}=\sum_1^{q}C^p_q\bigf_p\>\bigf_{q-p}{\gamma_0\over
p}\equat{II-7}
$$
where one can notice the transformation of the convolution in
\Form{II-5}
into a mere product, independent of the values of the
constants $c_p$.

 As a matter of fact, this set of equations is
identical to the one obtained for a generating
function of the global multiplicity distribution ${\cal H}(\Theta,u)$ of a
semi-random fragmentation model\ref{22-23} (Figure 1) which obeys the generic
equation:
$$
\dot{\cal H}={\cal H}(\widetilde{\cal H}-1)\equat{II-8}
$$
with
$$\
\widetilde{\cal H}(u)=\int_0^1 r({\rm w}){\cal H}(u{\rm w})d{\rm
w}\equat{II-9}
$$
whose            kernel is given by the distribution:
$$
r({\rm w})=\delta({\rm w})+{\gamma_0\over{\rm
w}}_{\big|_+}\equat{II-10}
$$
where the + stands for the Principal-Value distribution. The equivalence
between \Form{II-7} and \Form{II-8} is easily obtained by differentiation
in $u$ at $u=0$.
\midinsert
\figur{fig1.ps}
%\fig{12}{fig1.ps}
\caption{{\it Figure 1} : Sketch of the random fragmentation cascade. In grey,
 the "history" of the bin $\Delta$, i.e. the successive random multiplicative
factors ${\rm w}$ which define the multiplicity density in $\Delta$ for
one event. $Y$ (resp. $\Delta$) is the initial (resp. final) observation range.
As
a simple illustration from the figure, the bin density is $\rho_{\Delta}=
\rho_{Y}
 {\rm w}_{1}{\rm w}_{2}{\rm w}_{3}   {\rm w}_{4} $ where $\rho_{Y}$ is the
initial density,
and the w's are the random
or unity weights (see text)
corresponding to the semi-random structure.}
\bigskip
\endinsert

More precisely, the equations \Form{II-3}, \Form{II-9} are
obtained in the framework of Random-Branching Random-Cascading
fragmentation models\ref{22-23}, with a "time" variable to be identified, up
to
a constant,  with
$\ln \Theta_0^{-1}$. The model can be described as a tree of
random
multiplicative weights w (Figure 1), with the following recepe:
There is a unit probability per unit time to have a new
branching; At each such vertex, one branch brings the (deterministic)
weight 1 and the other gives a random weight w following the
probability law $r$(w) defined by \Form{II-9}. In \ref{22}, it has been
called a
 {\it semi-random} model.
 The $q\neq 0$ moments
of the distribution $r$ are precisely $\overline{\rm
w^q}=\gamma_0/q$, while its normalization is kept to be $1$.
This kind of kernel has been used in another context\ref{22} and
exhibits explicit intermittency properties . For this class
of fragmentation models to be compared with QCD
cascading,
the corresponding (positive) time variable runs from $0$ to
$\ln E \Theta_0/\mu$.

Now comes the step between the {\it global} formulation of a
random cascading model (formula \Form{II-7} for the
multiplicity moments $\bigf_q$ and \Form{II-8} for their generating
functions ${\cal H}$) to a {\it local} formulation for the factorial
moments ${\cal F}_q$ defined in \Form{I-1}. For this sake, one used to
introduce the generating fonction of the local moments, namely $H_\Delta$.

In the random fragmentation models described in
ref. [22-23], the development of the cascade gives rise
to the following equation governing the {\it local}
intermittent fluctuations:
$$
\dot{H}_\Delta=\widetilde{H}_\Delta - H_\Delta,\equat{II-11}
$$
where $H_\Delta$ is thus the generating function of the multiplicity
distribution
in a cell of size $\Delta$.
Qualitatively, Equation \Form{II-11} comes from the fact that along the path
leading to
the cell $\Delta$, see Figure 1, one encounters either weights $1$ or w, taking
 into account the semi-random structure of the model. The technical derivation
requires some care and has been obtained in two ways, either from the global
equation \ref {22}  or from a subdimensional equation \ref {23}. Note
that in order to figure out the (average) number of cascading steps,
the size of the cell $\Delta$ is to be compared with the large
scale of the problem, namely the initial range $Y$ (this model was originally
used for rapidity cells\ref{22,23}, but it can be extended to any other
relevant variable). The
corresponding "time" variable is indeed related to $t=\log {Y/\Delta}$.

Let us now come back to QCD. Indeed, in the present case, the cell one
is led to consider is {\it angular}, and its size $\Theta$ has to be
compared with the only large scale one has in
the problem (on an event by event basis),
namely the observation angle $\Theta_0$. Thus, the natural evolution variable
 involved in the
equation \Form{II-11} is $t=\log\Theta_0/\Theta$. This choice  can also be
understood as a reflexion of the QCD angular ordering \ref{20};
Angular ordering is the property of QCD jets at leading-log order
that, when partons are detected in the direction $\Theta_0$
with respect to a jet axis, the
maximal aperture of the "sub-jet" of partons contributing to the observation
is of the order of the observation
angle, with a subsequent degradation of the diffusion angles along the cascade.

{}From \Form{II-11}, one obtains  the following
solution for the angular factorial moments :
$$
\eqalign{{\cal F}_q(\Theta_0,\Theta)&\propto
\left[{\Theta_0\over\Theta}\right]^{(q-1)(1-{\cal D}_q)},\>{\rm with}
\cr
{\cal D}_q&={\gamma_0}{q+1\over q}\cr}\equat{II-12}
$$
expressing an exact power law \Form{I-2} in the angular variable
$\Theta$,
for a
fixed coupling constant, i.e. fixed $\g0$.

A series of remarks can be made about the result
\Form{II-12}.

\hskip 1truecm{\it i) The fixed coupling
constant regime}

 The intermittency property \Form{II-12} is
exactly realized at fixed coupling constant. However, one has to
verify the condition ${\cal D}_{q}<1$.
or, in words, that the Renyi dimension is smaller than the
support dimension of the
set of fluctuations\ref{5,3}. In \Form{II-12}
this is obtained for $\g0\leq \frac{2}{3}$,( $q\geq2$), or
equivalently, $\as/2\pi\leq 1/9N_c$, which means a rather weak
coupling regime. This corresponds to the QCD parton cascade in its
initial development. A correct treatment of the full cascade requires
taking into account the running QCD coupling.

\hskip 1truecm{\it ii) 2-dimensional result and collimation}

Motivated by the appearance of an angular
variable in \Form {II-12}, one can introduce the azimuthal
angle $\Phi$ in the game by considering the jet development in the
overall solid angle specified by the couple ($\Theta, \Phi$).
 Indeed, a slight modification of the arguments developped for \Form{II-11}
shows that one can embed a two-dimensional branching in a
higher $d$-dimensional phase space; It is enough (with some conditions of
uniformity\ref{24}) to introduce
($d-1$) other branches at the same vertex with zero multiplicative
weight (no particle produced), in the definition of the random
cascading model. This
can be interpreted as the {\it
collimation}
property of QCD jets.
 Indeed, it is not too difficult to realize that the {\it global} QCD
equation  \Form{II-3} remains unchanged by integration over the $d-1$
other
variables ($\Phi$, in our case) while the {\it local}
equation \Form{II-11} becomes :
$$
\dot{H}_\Delta=\widetilde{H}_\Delta-dH_\Delta+d-1\equat{II-13}
$$
With this physical interpretation and using \Form{II-13}, one finds
for the embedding into the 2-dimensional space ($\Theta, \Phi$),
 i.e for the  factorial moments
 in a solid angle phase-space cell of size $\Delta\Omega\simeq \pi
\Theta^2$,
$$
\eqalign{{\cal F}_q(\Delta\Omega)&\propto
\left[{\Theta_0\over\Theta}\right]^{2(q-1)(1-{\cal D}
_q)},\>{\rm with}
\cr
&{\cal D}_q={\gamma_0\over 2}{q+1\over q}\cr}\equat{II-14}
$$
Note that the
energy-momentum development of the QCD cascade corresponding to
formula \Form{II-14}
implies a "2/3-collimation" at each branching, since 1/3 of the cone
contains no particles in the equivalent fragmentation-model
framework.

\hskip 1truecm{\it iii) The running coupling constant case}

The extension of the fluctuation analysis to
the QCD running coupling constant is not too difficult. The
whole derivation essentially remains the same, except that one has to
take into account the $\as$ dependence on $\Theta$ in the solution of
the differential equations \Form{II-3} to  \Form{II-11}.
The final formula reads:
$$
\eqalign{{\cal F}_q(\Theta_0,\Theta)&\propto
\left[{\Theta_0\over\Theta}\right]^{d(q-1)(1-{\cal
D}_q)},\>{\rm with}
\cr
&{\cal D}_q(\Theta,\Theta_0)
={q+1\over dq}{\int_\Theta^{\Theta_0}
\gamma_0(\as(\theta))\>d\theta/\theta\over\ln\Theta_0/\Theta}\cr}
\equat{II-15}
$$
One retrieves expression \Form{II-12} by the approximation
$\as=cst$. As can be remarked from formula \Form{II-15}, the $\ln
\Theta$ variation of the coupling constant will induce a modification
of the behaviour of ${\cal D}_q$ which will, in general, depends on
both $\Theta_0$ and $\Theta$ and not on their only ratio. Thus one
may expect a modification of
the power-law \Form{II-14} - and of the intermittent behaviour
\Form{I-2} - due to the running of the coupling constant.
This effect is in fact related to the well-known QCD scaling
violation
of the structure functions observed in deep inelastic scattering.
 It is thus of importance to take into account the correct
$\as(\Theta)$ behaviour to compute the Renyi dimensions
${\cal D}_q$ in \Form{II-15}. However, this behaviour
depends on the genuine scale    on which $\as$ depends.
 A good candidate\ref{21}
is the parton relative transverse momentum $P_\perp$, but then, care
must be taken of the evolution of the parton energy $k$ in the jet,
since $P_\perp\simeq k\Theta$. Neglecting the energy loss      would
allow us to choose $k\equiv E$ (the initial energy of the gluon
jet), another extreme being to consider $k\equiv \mu$, the infra-red
cut-off;
We shall complete the discussion in the next section III,
removing this ambiguity, and shall define the concept of {\it angular
intermittency}
implied by \Form {II-15}.
\bigskip
{\bf III Intermittency from correlations in a QCD jet}
\bigskip
Let us now introduce the method of refs. \ref{18,19}.
For sake of simplicity, we start  again considering the fixed
coupling regime.
The correlation of two multiplicity flows in this case is sketched
in the
graphs of Fig.2a-b.
Here the gluon jet with initial energy $E$ and production angle
$\Theta_0$
evolves producing a soft offspring $k$ which then splits into two
partons
(gluons) with relative angle $\Theta_{12}\leq \Theta\ll \Theta_0$
generating the registered
particle flows.
Due to angular ordering the resulting 2-body correlation can be
written (after a few thechnical manipulations described farther) as a
convolution of the energy spectrum of parton $k$ with the product of
the two
multiplicity factors, namely :
$$
(4\pi)^2(1-\cos\Theta_{12})\frac{d^2 N^{(2)}}{d\Omega_1\,
d\Omega_2}=4C_F\frac{\as}{2\pi}
\int^{E}\frac{dk}{k}\>
\tilde{D}\left(\frac{E}{k},\frac{\Theta_0}{\Theta}\right)\cdot
2\dot{N}(k\Theta_{12})N(k\Theta_{12})\equat{III-1}
$$
where $\tilde{D}$ denotes the energy spectrum originating from
cascades
with parton emission angles {\it larger} than the given $\Theta$ and
$\dot{N}$ means the multiplicity derivative with respect to the
argument.
Notice that the colour factor $C_F$ in the emission probability of
the gluon $E$
corresponds to the case of the quark as an original parton shown by a
horizontal
line in Figs.2. This
quark line can be thought of as determining the main direction
of the hard
process under consideration (e.g., the quark jet direction in
$e^+e^-$ annihilation).
Note that Eq.\Form{III-1} is written for the  differential
distribution as a
function of the solid angles $\Omega_1$ and $\Omega_2$,
with $\Theta_{12}$ the angle between the two triggered partons.
\midinsert
\figur{fig2.ps}
%\fig{10}{fig2.ps}
\caption{{\it Figure 2} : Kinematics of parton-parton QCD
correlations;
a) Location   of the phase-space cell $\Delta$ in polar coordinates;
b)
schematic
representation of the convolution formulae [ III - 1,5 ].}
\bigskip
\endinsert
    The last factor in \Form{III-1} has emerged after performing the
integration
over the relative energy fraction, $z$, in the decay of the parton
$k$,
$$
\int_0^1 dz\>\> 4C_A\frac{\as}{2\pi}
\left\{\frac{1}{z}+\frac{1}{1-z}+\ldots\right\}
\>N(zk\Theta_{12})N((1\!-\!z)k\Theta_{12}) \approx 2
\dot{N}(k\Theta_{12})N(k\Theta_{12})\equat{III-2}
$$
where we only accounted for the infrared-singular (logarithmic) terms
of
the gluon
splitting probability and made use of the DLLA evolution equation
for the
(gluon) jet multiplicity\ref{20},
$$
\dot{N}(Q)\equiv \partder{\ln Q} N(Q) = \int_0^1 \frac{dz}{z}\ \
4C_A\frac{\as}{2\pi} \,N(zQ)\>\equat{III-3}
$$

To obtain the factorial moment $F_2$, i.e. the {\it integral}
multiplicity correlator, one has to integrate \Form{III-1} over
$d\Omega_1d\Omega_2$ keeping the relative angle between the two
directions smaller
than $\Theta$.  Observing that
$$
\int \frac{d\Omega_1d\Omega_2}{(4\pi)^2}
\frac{\vartheta\left(\Theta\!-\!\Theta_{12}\right)}{(1-\cos\Theta_{12}
)}
= \frac{d\Omega_1}{4\pi} \int_0^{2\pi} \frac{d\phi_{12}}{2\pi}
\int_0^{\Theta} \frac{\sin\Theta_{12}\>d
\Theta_{12}}{2(1-\cos\Theta_{12})}
\approx \frac{d\Omega_1}{4\pi} \int_0^{\Theta}
\frac{d\Theta_{12}}{\Theta_{12}}\equat{III-4}
$$
we obtain the formula :
$$
(4\pi)\frac{\Delta N^{(2)}}{\Delta\Omega}=\frac{4C_f\as}{2\pi}
\int^{E}\frac{dk}{k}\>
\tilde{D}\left(\frac{E}{k},\frac{\Theta_0}{\Theta}\right)
\cdot N^2(k\Theta)\equat{III-5}
$$
where $\Delta\Omega$ stands for the (small) solid angle size of the
phase space cell in consideration (see Fig. 2b).

The evolution equation for the distribution $\tilde{D}$,
$$
\tilde{D}\left(\frac{E}{k},\frac{\Theta_0}{\Theta}\right)=
\delta(\ln E/k)+
\int^{\Theta_0}_{\Theta}\frac{d\Theta'}{\Theta'}
\int_k^E\frac{d\ell}{\ell}\>4C_A\frac{\as}{2\pi}
\tilde{D}\left(\frac{E}{\ell},\frac{\Theta_0}{\Theta'}\right)\>,\equat
{III-6}
 $$
 can be solved by means of a Mellin transform in the $k/E$ variable
which gives :

 $$
 \tilde{D}_n({\Theta_0\over\Theta})=\exp{-{\gamma_0^2\over n}
\ln\left({\Theta\over\Theta_0}\right)}\equat{III-7}
 $$
 where
$$
\tilde{D}_n({\Theta_0\over\Theta})=\int_0^1\>x^{(n-1)}\>\tilde{D}\left
(
{1\over x},\frac{\Theta_0}{\Theta}\right)dx,\ x\equiv k/E
$$
 Using now the stationnary phase argument in the inverse Mellin
transform one obtains:
 $$
 \tilde{D}\left(\frac{E}{k},\frac{\Theta_0}{\Theta}\right)\simeq\exp{2
\gamma_0\sqrt{\ln\frac{E}{k}\ln\frac{\Theta_0}{\Theta}}}\equat{III-8}
 $$
 $\tilde{D}$ monotonically increases while $k$ decreases  down to the
value $\mu/\Theta$ where
$\mu$ denotes again the infra-red cutoff.
Starting from \Form{III-5}, we need to convolute the $\tilde{D}$
distribution with the  square
of the multiplicity factor .

More generally,  when not only two but $q>2$ particles are registered
in the final
state within the small cell $\Theta$, one has to consider
  the $q^{th}$ inclusive multiplicity correlator . Using
similar arguments, related to the DLLA properties of
Perturbative QCD for a jet\ref{18}, one is led to weight $\tilde{D}$
with the $q^{th}$ power of the mean multiplicity. This gives :
$$
(4\pi){\Delta N^{(q)}\over \Delta \Omega
}=\frac{4C_f\as}{2\pi}\int^E{dk\over k}\int {dn\over
2i\pi}\exp{S(n,k,\Theta)}\equat{III-9}
$$
with

$$
S(n,k,\Theta)=n\ln({E\over k})+{\g0^2\over n}
\ln({\Theta_0\over\Theta})+q\g0\ln({k\Theta\over\mu})\equat{III-10}
$$
where the introduction of the infra-red cut-off $\mu$ comes with the
multiplicity, see \Form{II-4}.
We can now make use of the steepest descent technics in $k$ together
with the
stationnary phase space approximation in $n$ and get    the result:
$$
\eqalign{(4\pi){\Delta N^{(q)}\over \Delta \Omega }&\propto
\left({E\Theta\over\mu}\right)^{q\g0}\left({\Theta_0\over
\Theta}\right)^{\g0\over q}\>,\>{\rm with}\cr
\ln{E\over k}&={1\over q^2}\ln{\Theta_0\over\Theta}\cr}\equat{III-11}
$$
Note that, in this case of a fixed coupling constant, there may be a
sizeable difference between the hierarchy in angles and the one
in energy,
at least for small-rank correlations.
The result \Form{III-10} is valid if one remains in the
perturbative regime i.e. $\ln\left({k\Theta\over\mu}\right)>0$.
Defining the rescaled variable $x_\mu$ as:
$$
x_\mu={\ln\left({\Theta_0\over\Theta}\right)\over\ln\left({E\Theta_0\over
\mu}\right)}\equat{III-12}
$$
where $\mu/E $ is the minimal angle $\Theta$ one may safely consider,
the previous
 condition reads:

$$
x_\mu<{q^2\over q^2+1},\>(={4\over5}\>{\rm when}\ q=2\>{\rm and}\
\simeq
1\>{\rm for}\>q\ {\rm large}).
$$
$x_\mu >{4\over 5}$ is thus the correct limit of the
perturbative regime.

The scaled moments can now be calculated, and using
$\Delta\Omega\simeq \pi\Theta^2$, one retrieves formula \Form{II-14}
under the
guise:
$$
\eqalign{{\cal F}_q(\Delta\Omega)&\propto
\left[{\Theta_0^2\over\Theta^2}\right]^{(q-1)(1-{\cal D}_q)},\>{\rm
with}
\cr
{\cal D}_q&={\gamma_0\over 2}{q+1\over q}\cr}\equat{III-13}
$$
and, by integration on the azimuth, formula II-12.

Let us come back now to the perturbative QCD calculation with the
appropriate running coupling constant $\as(P_{\perp}=k\Theta)$.
To estimate the typical momenta $k$ in the main convolution formula
\Form{III-1}
we use again the Mellin transform, as in formulae \Form{III-9,-10},
but in the context of a running $\as$, one gets :
$$
\frac{\Delta N^{(q)}}{\Delta \Omega} \propto \int\frac{dk}{k}
\int\frac{d\omega}{2\pi i} \>
\exp{{\cal{S}}(\omega, k,\Theta)}\equat{III-14}
$$
with
$$
{\cal{S}}(\omega,k,\Theta)= \omega\ln\frac{E}{k}+
\int_{k\Theta}^{E\Theta_0} \frac{dt}t\>\gamma_\omega(\as(t)) +
 q\cdot\int_{\mu}^{k\Theta}\frac{dt}t\>\g0(\as(t))\equat{III-15}
$$
and
$$
\gamma_\omega(\as)=\half
\left[\>-\omega+\sqrt{\omega^2+4\g0^2}\>\right]\>\equat{III-16}
$$
where $\gamma_\omega$ is  the standard DLA anomalous dimension.
Then we evaluate [III-14] by steepest descent in $k$ and stationary
phase
in $\omega$ to get
$$
\eqalign{-&\partder{\ln k} {\cal{S}}=\omega + \gamma_\omega(\as(k\Theta))
-
q\g0(\as(k\Theta)) = 0 \cr
&\partder{\omega} {\cal{S}} =\ln\frac{E}k
+ \int_{k\Theta}^{E\Theta_0}\frac{dt}{t}\partder{\omega}
\>\gamma_\omega(\as(t)) = 0\cr}\equat{III-17}
$$
Making use of \Form{III-15} we find the following solution for the
preferred value $\moy{\omega}$ :
$$
\moy{\omega} = \frac{q^2-1}{q}\>\g0(\as(k\Theta))\equat{III-18}
$$
and substitute it into the second relation \Form{III-16} to obtain
the
equation which determines implicitly $\moy{k}$:
$$
{\ln\frac{E}{k}} = \ln\frac{\Theta_0}{\Theta}
 -\int_{k\Theta}^{E\Theta_0}\frac{dt}{t}\frac1{\sqrt{1+\left[\frac{2q}
{q^2-1}
\right]^2\frac{\as(t)}{\as(k\Theta)}}}\equat{III-19}
$$

While     in the fixed coupling regime we recover all the previous
results, for
example :
$$
{\ln\frac Ek} = \frac{1}{q^2} \ln\frac{\Theta_0}{\Theta}
\left( =\frac{1}{4}
 \ln\frac{\Theta_0}{\Theta} \>\>{{\rm for}\>
q=2}\>\right)\>.\equat{III-20}
$$
we get now:
 $$
\eqalign{&
x=\half\left[1-y\left(1-\int_{y}^1\frac{du}{u^2}\frac{1}{\sqrt{1+a_q
u}}\right)\right],\>a_q=\left(\frac{2q}{q^2-1}\right)^2\cr
&x={\ln \Theta_0 / \Theta \over \ln E \Theta_0 /
\Lambda} \>;\>\>\>y=\frac{\ln k \Theta /\Lambda }{\ln
E\Theta_0/\Lambda},\cr}\equat{III-21}
$$
where $\Lambda $ is the QCD scale at first perturbation order.

It is important to notice
that the QCD scale $\Lambda$ has replaced the infrared cut-off $\mu$
which appeared in the variable $x_\mu$, see
\Form{III-12}. In fact, $\mu$ appears only as the
lower bound of the multiplicity contribution to ${\cal
S}(\omega,k,\Theta)$, see formula \Form{III-15}, and
disappears from the {\it scaled} factorial moments.
The final formula will be independent of the infra-red cut-off.
\midinsert
\figur{fig3.ps}
%\fig{14}{fig3.ps}
\caption{{\it Figure 3} : Comparison of angular/momentum hierarchies
in a QCD jet fragmentation;
a)  The scaled-angular $x$ variable as a function
of the transverse momentum one $y$.
continuous line : The exact formula [ III - 21 ] ;
dotted-dashed line : $x=1-y$ corresponding to  $k=E$. b) The
QCD second intermittency dimension as a function of $\tilde{x}$ ;
continuous
line : the exact [ III - 23 ] formula; dashed curve : the
approximate
formula [ III - 22 ].}
\medskip
\endinsert
The function $ x(y)$ determines how the angular hierarchy in
$\ln\Theta_0/\Theta $ - measured by its ratio with the maximal
value     $\ln E \Theta_0/\Lambda$ - depends on the transverse
momentum hierarchy  expressed    by $1-y={\ln P_
{\perp_0}/P_\perp\over \ln P_{\perp_0}/\Lambda}$.
The function $x(y)$ is displayed in figure 3 for $q=2$, together with
the function $1-y$, which would correspond to taking
$\as(k\Theta)=\as(E\Theta)$.
    The lesson from Figure 3a is  that it is quite
consistent to chose $k\simeq E$ in the angular evolution of the
running coupling constant $\as(k\Theta)$.
The function $x$ can be choosen equal to $1-y$ with a quite good
approximation. This leads to the following simple expression fo
${\cal D}_q$ (cf. also formula \Form{II-15})
$$
\eqalign{{\cal D}_q&={q+1\over
dq}\frac{\int_\Theta^{\Theta_0}\g0(\as(E\theta))d\theta/\theta}
{\ln\Theta_0/\Theta}\ =\cr
&={q+1\over dq}\g0(\as(E\Theta_0)){2\over
x}(1-\sqrt{1-x}),\cr}\equat{III-22}
$$
where $x$ is the same as in \Form{III-21}, and d the dimension.
 Note again that
the result \Form{III-22} is independent of the infra-red cut-off.

To be complete, one can work out the exact calculation of the
 anomalous dimension as
a function of $1-y=\tilde{x}$:
$$
\eqalign{
{\cal D}_q(\tilde{x})
&=2\g0{{q+1}\over dq}\cr
&\frac{(1-\sqrt{1-\tilde{x}})q\sqrt{a_q}-a_q\sqrt{1-\tilde{x}}\,\ln\left[
q^2\frac{1-\sqrt{1+a_q(1-\tilde{x})}}{1+\sqrt{1+a_q(1-\tilde{x})}}\right]
}{1-(1-\tilde{x})q\sqrt{a_q}+\sqrt{1+a_q(1-\tilde{x})}+a_q(1-\tilde
{x})\ln\left[q^2\frac{1-\sqrt{1+a_q(1-\tilde{x})}}{1+\sqrt{1+a_q(1-\tilde
{x})}}\right]}\cr
&\simeq 2\g0{q+1\over
dq}\left({1-\sqrt{1-\tilde{x}}\over
\tilde{x}}\right)\cr
}\equat{III-23}
$$
The exact anomalous dimension ${\cal D}_2$ is displayed in Figure 3b
for $d=2$, together with the approximate one, as a function of
$\tilde{x}$. The figure shows that the  formula \Form{III-21}
(also
\Form{II-15}) is
very close to the exact one. This ensures  the equivalence between
the fluctuation and correlation approaches
to the intermittency calculations.
In other terms, for factorial moments,
the hierarchy in cascading angles is more stringent than
the softening in energy. Indeed, both curves differ only slightly in
the region
$\tilde{x}\simeq 1$ which can be considered outside the validity
region for a perturbative calculation, cf. the discussion in the
next section IV.
\bigskip
{\bf IV Intermittency properties of a QCD jet}
\bigskip
Let us now examine in a more precise way the intermittency
properties of the jet, as given by the final formula
 \Form{III-22}.
We have already noticed that in the case of a fixed running constant,
one verifies an exact scaling law \Form{I-2}, where the phase-space
cell $\Delta$,  in one dimension, is angular. We shall now
discuss the physical meaning of the result obtained for the running
coupling constant in case of 1 and 2 dimensions. Let us pinpoint a
few
properties.
\bigskip
{\it Angular intermittency of QCD jets}

The main feature of formula \Form{III-22} is that it is expressed
in terms of angular variables, namely the diffusion
angle $\Theta$ ( in 1 dimension) and the solid angle $\Omega$ (in 2
dimensions). In terms of $\Theta$ in particular, or, more precisely,
in terms of the variable $x=\frac{\ln\Theta_0/\Theta}{\ln
E\Theta_0/\Lambda}$, a quite "universal" behaviour is observed, with
increasing factorial moments at small $x$, and then saturation and
fall near $x=1$ (see Fig. 4a for 1 dimension, Fig. 4b for 2
dimensions).
This behaviour can be easily understood from \Form{III-22}. At small
values of $x$ - or not too small cell-sizes $\Theta$ -
the intermittent power-law
\Form{II-12,14} is approximatively valid provided one considers the
coupling
constant $\g0$ at $k=E$, i.e. its minimal value within the jet.
Then,
for smaller cell-size, an
interesting  {\it saturation effect} occurs, which is obtained in the
framework of perturbative QCD, independently of any
hadronisation model. This {\it saturation effect }  is entirely due to
the running of the coupling constant through   the increase of
the Renyi dimension with $x$, see \Form{III-22}. Indeed, this
increase becomes so strong that there is an oversaturation effect in
the theory, at a value of $\g0$ of the order of 1, which sets the
limit of validity of the perturbative expansion of the theory. The physical
interpretation is simple, the many gluons produced saturate the fluctuations;
indeed, there are so many, that one can no more neglect there reinteractions :
We are entering the non perturbative regime.

 In Fig. 4b, the result for the two-dimensional
case is shown and we remark that the saturation effects are milder;
This is a reflexion of the dividing factor $d$ in the
Renyi dimension which increases the intermittency strength
and thus decreases the relative
importance of the saturation effect.
\midinsert
\figur{fig4.ps}
%\fig{14}{fig4.ps}
\caption{{\it Figure 4} : QCD predictions for the first moments
${\cal F}_q$
as a function of $x$; a)
$d=1$, b) $d=2$. The predictions are computed for the following
parameter value :
$\epsilon\simeq 5$, which for a jet energy $E\simeq 50 Gev$ and
$\Lambda=200$ Mev, say, corresponds to an angular aperture of
 $\Theta_0\simeq .6$.}
\bigskip
\endinsert
What is the implication of angular intermittency for the
usual analysis framework
\ref{1,8-12} ? The natural scaling variables emerging from
\Form{III-21}) are $x$ and $\epsilon$, which are related to angles
by:
$$
\ln
\Theta_0/\Theta=x.\epsilon  = x.\>\ln(E\Theta_0/\Lambda),\equat{IV-1}
$$
where the scaling parameter $\epsilon=\ln(E\Theta_0/\Lambda)$
contains all the information on the ki\-ne\-ma\-tical parameters of
the considered jet evolution (Note that, from Eq. \Form{III-22},
${\cal D}_q$ depends on $x$ {\it and} on $\g0(\as(E\Theta_0))$,
which  itself depends   of $\epsilon$). One important
consequence is that the value of $\epsilon$ fixes both the energy
$(E)$
and the angular $(\Theta_0)$ dependence of the fluctuation patterns.
 An example of this dependence on $\epsilon$ is given in Fig. 5,
 for
${\cal F}_2$ in two dimensions for 2 typical values of $\epsilon$.
\midinsert
\figur{fig5.ps}
%\fig{10}{fig5.ps}
\caption{{\it Figure 5} : 2-dimensional ${\cal F}_2$ as a function
of
$\ln\Theta_0/\Theta$ for two
different values of $\epsilon$ (see text).}
\bigskip
\endinsert
Note that the saturation effect is enhanced at smaller
$\epsilon$ which means smaller energy and/or smaller angle. We will
come back to this interesting property.
\bigskip
{\it Comparison with previous approaches}

It is instructive to confront the obtained results -
\Form{III-22,23} - and the two different methods we used,
fluctuations and correlations, with previous works on the
subject\ref{12-17}.
The formulae \Form{III-22,23} and the
method leading to them are worth comparing with the results of
ref.[15-17]. In ref.[15], the dipole approximation of perturbative
QCD cascading led to a determination of the fractal
dimension of the so-called "$\lambda$ curve" at large values of $q$.
This
dimension coincide with the large-$q$ limit of the approximate Renyi
dimension \Form{III-22}. On the other hand, the small-$q$ evolution
of the Renyi dimension which characterises the "multi-fractal"
feature
of QCD gluon cascading is numerically evaluated.
Note however, a clear difference in the interpretation of the result:
The conclusion of \ref{15} is to propose observables different from
 factorial
moments to exhibit the scale invariant properties of gluon cascading,
attributing to hadronisation a dynamical role in the fluctuation
pattern.
At the level of hadrons, the question is      whether  gluon-hadron
duality is valid or whether hadronisation changes the features of
factorial moment observables as suggested in [15]. This deserves
more
investigation.

We have borrowed from \ref{16} the method using the non-linear
equation
for multiplicity moments and its transformation in a fragmentation
formulation.
 However, Ref.\ref{16}         was only considering planar diagrams
with infra-red singular contributions, which are cancelled in the
full DLLA (planar + non-planar) diagrams.
A comparison with ref.[17] shows that
 the
occurence of a different dominant singularity at small angles, found
in
\ref{17}, is not found in our analysis. Our results are in agreement
with
those found in ref.[18].

\bigskip
{\it Phenomenological discussion}

Some features present in the Monte-Carlo
estimates\ref{11} of the factorial moments based on a Perturbative
QCD ansatz are explicit in our analytical solutions, for example  the
saturation
at small cell-size and collimation in azimuth. However, a detailed
phenomenological discussion requires an hypothesis on the
transformation of partons
into hadrons. In fact, we will
consider that
the effect of hadronisation -which is model-dependent in the
Monte-Carlo's-
 is  smooth enough to allow for a
direct comparison of hadron experimental data with the predictions
of  formulae \Form{III-22,23}. This  smoothness was already
noticed for other observables such as the one-particle inclusive
cross-section, the well-known humped-backed plateau in \epm
annihilation into hadrons, providing an argument in favor
of the LPHD assumption\ref{14}. It is interesting to
discuss whether or not this
property may also apply to multiparticle correlations/fluctuations
effects, using factorial moment observables. Note, however, that we
did not
enter in the computation of next-to-leading orders, which are known
\ref{19,20}
to be important in 2-particle or higher order correlations. Indeed,
DLLA
partonic evaluation seems to overestimates the correlations in a
systematic
way. We thus have to find predictions which would minimize the
higher-order
 and hadronisation corrections.

The experimental data on factorial moments and intermittency in \epm
anni\-hilation into hadrons and espe\-cially on hadronic decays of
${\cal Z}_0$'s are al\-ready remar\-kable\ref{10,11}. A few
features which we discussed previously, like the saturation effects,
the increase of intermittency with dimensionality, the
dependence of the diffusion direction angle can be seen in the data.
However, the tests of angular intermittency need a reappraisal
of the most convenient observables. Indeed, no data exist
directly in terms of the $\Theta$ variables, and it
would
be unfair to make uncontrolled combinations of
 the existing data \ref{11} depending on rapidity. We thus propose
new observables in order to check the QCD predictions.

The best indication for confronting angular intermittency with data
- if experimentally feasible - would be to give
one- and two-dimensional factorial moments directly for small
cells in angular variables. Indeed, let us
consider a given direction $\Theta_0$, and two angular cells
of aperture $\Theta_1$ and $\Theta_2$ around that direction. Using
twice
\Form{II-15}, one gets:
$$
{\cal F}_q(\Theta_1)={\cal F}_q(\Theta_2)\ \exp{(q-1)\left
(\log\Theta_2/
\Theta_1
-\frac{q+1}{q}\int_{\Theta_1}^{\Theta_2}\g0(\as(E\theta))d\theta/
\theta\right )}\equat{IV-2}
$$
\Form{IV-2} is formally independent of the chosen
direction $\Theta _0$, except that one has to take into account
 the application range of perturbative
calculations, namely:
$$
0<\log \frac{\mu}{\Lambda}<\log \frac{E\Theta_1}{\Lambda}<
\log \frac{E\Theta_2}{\Lambda}<\log
\frac{E\Theta_0}{\Lambda}=\epsilon
\equat{IV-3}
$$
For example, in Fig. 6, we
display the predictions for ${\cal Z}_0$ jet decays
 corresponding to 1-dimensional ($\varphi$-averaged) and
2-dimensional moments of rank 3 as a function of $\log{\Theta_2/\Theta_1}$
for two typical   values of $\epsilon_2=\log{E\Theta_2/\Lambda}$.
Following \Form{IV-3}, the range of the variable $\log{\Theta_2/\Theta_1}$
between
$0$ and $\log E\Theta_0/\mu \simeq 3$.
\midinsert
\figur{fig6.ps}
%\fig{14}{fig6.ps}
\caption{{\it Figure 6} : ${\cal F}_3$ {\it ratios}  as a function of
${\Theta_2/\Theta_1}$ (in log/log units);
a) $\epsilon_2=6$; b) $\epsilon_2=4$.}
\bigskip
\endinsert
Note that the saturation effect, clearly present in the one
dimensional case, almost desappears for 2-d solid angles cells.
Data on factorial moments are in general
 (except those from ALEPH collaboration)
"horizontally" averaged over different rapidities - i.e.
different $\Theta_0$ angles -. We suggest to perform {\it ratios} of
moments for different apertures before an eventual horizontal
averaging in order to test
the prediction \Form{IV-2}. It is interesting to notice that, if this
ratio
can be measured event by event, the result is predicted to be
independent
from the angular direction $\Theta _0$, and thus from the definition
of the
jet axis! However, one has to require to avoid small angles
$\Theta_0$,
(a  non-perturbative region for QCD) and thus the determination of
jets
cannot be totally ignored. In any case, the check of \Form{IV-2}
represents
a challenge for QCD phenomenology, taking into account that
higher-order
and hadronisation corrections are probably diminished in the ratio of
moments .

\bigskip
{\bf V Conclusions and outlooks}
\bigskip
The conclusions of our study can be cast in the form of a summary of
answers
to the questions $i-v)$ raised in the introduction;

$i)$ Perturbative QCD is {\it intrinsically intermittent}, at least
in the
Double Leading Logarithm Approximation (DLLA),
 since the factorial moments develop an (effective)
singular behaviour in small phase-space cells away from the jet axis,
see \Form
{II-15, III-22}. However, instead of a singularity in rapidity cell-size, as
in previous studies,
one obtains   a singularity in angular cell-size.

$ii)$ Intermittent patterns of fluctuations appear in  angular
variables,
namely diffusion and azimuthal angles around the jet axis, and solid
angle
for 2-dimensional fluctuations. The resulting {\it angular
intermittency}
manifests itself as the enhancement of  gluon  multiplicity
fluctuations in smaller and
smaller angular cells in a given direction away from the initial jet
axis.
Non-perturbative or higher-order corrections invalidate the
calculation for
too small cells or angles with respect to the jet axis.

$iii)$ The intermittency indices, characterizing the singular
behaviour of
fluctuations, are related to the QCD anomalous dimension of
multiplicities
$\g0$, and through it, depend only on the scaling parameter $\epsilon
=\log
{E\Theta/\Lambda}$, where $\Theta$ is the angular cell-size.

$iv)$ {\it Angular intermittency} predicts that the ratio of
factorial
moments for two cells of different size around the same direction
depends
only and specifically on these two sizes. We suggest to use angular
variables in the phenomenological analysis of $Z_0$ decays into
hadrons
to check this prediction, assuming parton-hadron
duality.

$v)$ The reason why the intermittency singularity appears is
primarily due
to the resummation of the infra-red potential divergencies of the
theory
through their leading-log contributions at all orders; The size of
the
angular     cell, over which one integrates to get factorial moments,
acts as
a regulator scale over these potential infra-red divergencies.

Questions to be solved in the near future are not lacking, starting
with
the abovementionned suggestions and requirements for experimental
studies.
  Among other problems, it
seems that the 3-d fluctuations, when one adds transverse momentum in
the
game, are at hand. The study of the transition to the
non-perturbative
region and its fluctuation properties could be facilitated by its
kinematical
location in the fragmentation rapidity domain. A related question is
to
ask whether other multi-particle reactions present
similar features as \epm . This would give an hint on possible mixing
of perturbative and other dynamical fluctuation mechanisms; One would
of course think to deep-inelastic reactions at HERA, but also to
hadron
or nuclei-induced reactions.

As a final and more speculative remark, it is tempting to consider
the QCD calculations as a new laboratory for the unsolved problem
of turbulence. It is quite clear that turbulence is by far a more
complex
problem, but the existence of a gauge field
theory like QCD supporting the
existence of self-similar fluctuation patterns and experiments of
high quality to discuss it could give some help.
After all, long ago, the possibility of a {\it fractal} behaviour
of jets into jets into jets... has been beautifully predicted
\ref{25}.
It is quite remarkable that the property of {\it angular
intermittency}
in QCD jets seems to realize this prediction in a quite non-trivial
way since there appears a $q$-dependent and angular-dependent set of
fractal dimensions (multi-fractality property).
\bigskip
{\noindent\bf Acknowledgments}
\bigskip
We warmly thank Yuri Dokshitzer for stimulating discussions which convinced
us to use for the present work one basic (correlation) method developed by him
and coworkers ; We acknowledge
him for helpful remarks and communicating related unpublished notes (with
Serguei Troyan).
We had also fruitful discussions with Bo Andersson, Andrzej Bialas, Wolfgang
Ochs and Jacek
Wosiek at the occasion of recent meetings and conferences (Orsay/Saclay,
Santiago, Lisbon,...). We thank Henri Navelet for final remarks on the draft.
\vfill\eject\

{\noindent\bf References}
\bigskip
{\leftskip=.7truecm

\noindent
\rm
\item{\bf 1} For a review and early references, see A. de Angelis, P.
Lipa
and W. Ochs, Proceedings of the Joint {\bf LP-HEP} Conference 91,
Vol. 1,
page 724 (Eds. S. Hegarty, K. Potter, E. Quercigh, World Scientific).

\noindent\item{\bf 2} A. Bia{\l}as and R. Peschanski, {\it Nucl.
Phys. B} {\bf 273} (1986) 703, {\bf 308} (1988) 857.

\noindent\item{\bf 3} For general review on the subject:
A. Bia{\l}as {\it Nucl. Phys. A}{\bf 525} 345c (1991); R.
Peschanski, {\it Int. J. Mod. Phys.}{\bf A6} (1991) 3681.

\noindent\item{\bf 4} P. Carruthers and I. Sarcevic, {\it Phys. Rev.
Lett.}
{\bf 63} (1989) 1562, and refs. in \ref{3}.

\noindent\item{\bf 5} P. Lipa and B. Buschbeck, {\it Phys. Lett.}{\bf
B223} (1989) 465 and references therein.

\noindent\item{\bf 6} P. Carruthers, {\it J. Stat. Phys.}{\bf
51} (1988) 517.

\noindent\item{\bf 7} A. Bia{\l}as and J. Seixas, {\it
Phys. Lett.}{\bf B250} (1990) 151.

\noindent\item{\bf 8} For a recent experimental review, see F.
Verbeure,
Proceedings of the XXII International Symposium on Multiparticle
Dynamics, Santiago de Compostela, 13-17 July 1992 (to be published
soon,
World Scientific). \par

\noindent\item{\bf 9} See, for instance, refs [1,3] and various
contributions
to the  proceedings
of ref. [8].

\noindent\item{\bf 10} B. Buschbeck, P. Lipa and R. Peschanski, {\it
Phys.
Lett.}{\bf B215} (1988) 788; HRS coll., TASSO coll., CELLO coll.
references
in [1].

\noindent\item{\bf 11} Last LEP results:
DELPHI Coll., P. Abreu et al. {\it Nucl. Phys.}{\bf B386} (1992) 471;
ALEPH Coll., D. Decamp et al. {\it Z. Phys.}{\bf C53} (1992) 21;
See previous references (including OPAL and L3 published
results, 1991) therein.

\noindent\item{\bf 12} M. Jedrejczak, {\sl Phys. Lett. B\/} {\bf 228
}(1988) 259. \par

\noindent\item{\bf 13} C.B. Chiu and R.C. Hwa, {\sl Phys. Lett. B\/}
{\bf 236
}(1990) 446. \par

\noindent\item{\bf 14} Y.I. Azimov, Y.L. Dokshitzer, V.A. Khoze, and
S.I.
Troyan, {\it Z. Phys.}{\bf C 31} (1986) 213. \par

\noindent\item{\bf 15} G. Gustafsson and A. Nilsson, {\it Z. Phys.}
{\bf C 52}
(1991) 533;

\noindent\item{\bf 16} Ph. Brax and R. Peschanski, Saclay Preprint
SphT/92-005(1992) .

\noindent\item{\bf 17}  W.
Ochs and J. Wosiek, {\sl Phys. Lett. B\/}{\bf 289} (1992) 159.

\noindent\item{\bf 18}
Y.L. Dokshitzer and S.I. Troyan, unpublished (1991).

\noindent\item{\bf 19} Y.L. Dokshitzer, G. Marchesini and G. Oriani,
{\it Nuclear
Physics }{\bf B 387} (1992) 675.

\noindent\item{\bf 20} For a review, see for example {\it Basics of
Perturbative QCD }  Y.L. Dokshitzer, V.A. Khoze, A.H. Mueller and S.I.
Troyan
(J. Tran Than Van ed., Editions Frontieres) 1991,
and the list of references therein.

\noindent\item{\bf 21} Note that the equation \Form{II - 2} is
valid beyond DLLA,                                          see ref.
[18], provided one chooses a running QCD scale $\as(P_\perp)$, where
$P_\perp\simeq k\Theta_0$, $k$ being the current parton energy with
an
infrared cut-off $\mu$; The condition : ${\cal
Z}(P_\perp\equiv\mu,v)=v$ has to be fulfiled.

\noindent\item{\bf 22}
J.-L. Meunier and R. Peschanski, {\it Nuclear
Physics }{\bf B 374} (1992) 327,

\noindent\item{\bf 23} Y. Gabellini, J.-L. Meunier and R.
Peschanski, {\it Z. F. Phys.} {\bf C 55} (1992) 455.

\noindent\item{\bf 24} W. Ochs, {\it Z. Phys.}{\bf C 50} (1991) 379;
{\it Act.
Phys. Pol.}{\bf B 25} (1991) 203.
A. Bialas and M. Gazdzicki, {\it Phys. Lett.}{\bf B 252} (1990) 483.

\noindent\item{\bf 25} G. Veneziano, Proc. 3rd. Workshop on Current
Problems in HEP theory, Florence 1979, eds. Casalbuoni et al;
John Hopkins University Press, Baltimore.
\par}

\vfill\eject
{\bf Figure captions}
\bigskip
{\it Figure 1} : Sketch of the random fragmentation cascade. In grey,
 the "history" of the bin $\Delta$, i.e. the successive random multiplicative
factors ${\rm w}$ which define the multiplicity density in $\Delta$ for
one event. $Y$ (resp. $\Delta$) is the initial (resp. final) observation range.
As
a simple illustration from the figure, the bin density is $\rho_{\Delta}=
\rho_{Y}
 {\rm w}_{1}{\rm w}_{2}{\rm w}_{3}   {\rm w}_{4} $ where $\rho_{Y}$ is the
initial density,
and the w's are the random
or unity weights (see text)
corresponding to the semi-random structure.
\bigskip
{\it Figure 2} : Kinematics of parton-parton QCD correlations;
a) Location   of the phase-space cell $\Delta$ in polar coordinates;
b)
Schematic
representation of the convolution formulae [ III - 1,5 ].
\bigskip

{\it Figure 3} : Comparison of angular/momentum hierarchies in a QCD
jet fragmentation;
a)  The scaled-angular $x$ variable as a function
of the transverse momentum one $y$.
Continuous line : The exact formula [ III - 21 ] ;
dotted-dashed line : $x=1-y$ corresponding to  $k=E$. b) The
QCD second intermittency dimension as a function of $\tilde{x}$ ;
continuous
line : the exact [ III - 23 ] formula; dashed curve : the approximate
formula [ III - 22 ].
\bigskip

{\it Figure 4} : QCD predictions for the first moments ${\cal F}_q$
as a function of $x$; a)
$d=1$, b) $d=2$. The predictions are computed for the following
parameter value :
$\epsilon\simeq 5$, which for a jet energy $E\simeq 50 Gev$ and
$\Lambda=200$ Mev, say, corresponds to an angular aperture of
 $\Theta_0\simeq .6$.
\bigskip

{\it Figure 5} : 2-dimensional ${\cal F}_2$ as a function of
$\ln\Theta_0/\Theta$ for two
different values of $\epsilon$ (see text).
\bigskip

{\it Figure 6} : ${\cal F}_3$ {\it ratios} as a function of $\Theta_2
/\Theta_1$ (in log/log units);
a) $\epsilon_2 = 6$; b) $\epsilon_2 = 4$.
\bigskip
\end